\documentclass{ws-procs961x669}            % default, citations in superscript

\newcommand{\rar}{\rightarrow}

\def\be{\begin{eqnarray}}
\def\ee{\end{eqnarray}}

\begin{document}
\title{Spontaneous and Gravitational Baryogenesis}

\author{E. V. Arbuzova$^*$ }

\address{ Department of Higher Mathematics, Dubna State University ,\\
Dubna, 141983, Moscow region, Russia\\
Physics Department, Novosibirsk State University,\\
Novosibirsk, 630090, Russia\\
$^*$E-mail: arbuzova@uni-dubna.ru\\
www.uni-dubna.ru, www.nsu.ru}

\begin{abstract}
Some problems of spontaneous and gravitational baryogenesis are discussed. Gravity modification due to the curvature dependent term in gravitational baryogensis scenario is considered. It is shown that the interaction of baryonic fields with the curvature scalar leads to strong instability of the gravitational 
equations of motion and as a result to noticeable distortion of the standard cosmology.
\end{abstract}


%\bodymatter

\section{Introduction}\label{aba:sec1}

Observations show that at least the region of the Universe around us is matter-dominated. Though we understand how the matter-antimatter asymmetry  may be created, the
concrete mechanism is yet unknown. The amount of antimatter is very 
small and it can be explained as the result of high energy collisions in space. The existence of large regions of antimatter in our neighbourhood would produce high energy radiation as a consequence of matter-antimatter annihilation, which is not observed. Any initial asymmetry at inflation could not solve the problem of observed excess of matter over antimatter, because the energy density associated with baryonic number would not allow for sufficiently long inflation. 

On the other hand, matter and antimatter seem to have  similar properties and therefore we could expect a matter-antimatter symmetric Universe. 
A satisfactory model of our Universe should be able to explain the origin of the local observed matter-antimatter asymmetry.
The term {\it baryogenesis} means the { generation of the asymmetry} between baryons 
(basically protons and neutrons) 
and antibaryons (antiprotons and antineutrons). 

In 1967 Andrey Sakharov pointed out 3 ingredients, today known as {\it Sakharov principles}, to produce a matter-antimatter asymmetry from an initially symmetric Universe. These conditions include: 1) non-conservation of baryonic number; 2) breaking of symmetry between particles and antiparticles;
3) deviation from thermal equilibrium. However, not all of three Sakharov principles are strictly necessary. 

In what follows we briefly discuss some features of spontaneous baryogenesis (SBG) and concentrate in more detail on gravitational baryogenesis (GBG).
Both these mechanisms do not demand an explicit C and CP violation and can proceed in thermal equilibrium. Moreover, they are usually  most efficient in 
thermal equilibrium. 

The statement that the cosmological baryon asymmetry can be created by spontaneous baryogenesis in thermal equilibrium was mentioned in the original 
paper by Cohen and Kaplan\cite{Cohen:1987vi} and developed in subsequent papers\cite{Cohen:1988kt,Cohen:1991iu}, for review 
see\cite{Dolgov:1991fr,Rubakov:1996vz}.

The term "spontaneous" is related to spontaneous breaking of a global $U(1)$-symmetry, which ensures the conservation of the total baryonic number in the 
unbroken phase. This symmetry is supposed to be spontaneously broken and in the broken phase the Lagrangian density acquires the additional term 
\be
{\cal L}_{SB} =  (\partial_{\mu} \theta) J^{\mu}_B\, ,
 \label{L-SB}
 \ee
 where $\theta$ is the Goldstone field and $J^{\mu}_B$ is the baryonic current of matter fields, which becomes non-conserved.
 
 For a spatially homogeneous field, $\theta = \theta(t)$, the Lagrangian is reduced to the simple form
  \be 
   {\cal L}_{SB} =   \dot \theta\, n_B\,, \ \ \ \   n_B\equiv J^0_B,
  \ee
  where time component of a current is the baryonic number density of  matter, so it is tempting to identify ${\dot \theta}$ with the chemical potential, 
  $ \mu_B$, of the corresponding system.  However, such identification 
%  of ${\dot\theta}$ with chemical potential 
  is questionable and
 depends upon the representation chosen for the fermionic fields\cite{Dolgov:1994zq,Dolgov:1996qq}.
It is heavily based on the assumption ${\dot \theta \approx const}$, which is relaxed in the work\cite{Arbuzova:2016qfh}.  
But still the scenario is operative and presents a beautiful possibility
to create an excess of particles over antiparticles in the Universe.

Subsequently the idea of gravitational baryogenesis (GBG) was put forward~\cite{Davoudiasl:2004gf}, where the 
scenario of SBG was modified by the introduction of the coupling of the baryonic current to the derivative 
of the curvature scalar $R$:
\be
{\cal L}_{GBG} = \frac{1}{M^2} (\partial_\mu  R ) J^\mu_B\, ,
\label{L-GBG}
\ee 
where $M$ is a constant parameter with the dimension of mass. 

In the presented talk we demonstrate that the addition of the curvature dependent term (\ref{L-GBG}) to the 
Hilbert-Einstein Lagrangian of General Relativity (GR) leads to higher order gravitational equations of motion, which are strongly unstable with
respect to small perturbations. The effects of this instability may drastically 
distort not only the usual cosmological history, but also the standard Newtonian gravitational dynamics. We discovered such instability for  scalar baryons~\cite{Arbuzova:2016cem}  and  found similar effect for the more usual spin one-half baryons (quarks)~\cite{Arbuzova:2017vdj}.

\section{Gravitational baryogenesis with scalar baryons \label{s-bosons}}

Let us start from the model where baryonic number is carried by scalar field $\phi$ with 
potential $U(\phi, \phi^* )$.
An example with baryonic current of fermions will be considered in the next section.

The action of the scalar model has the form:
\begin{eqnarray}
A = \int d^4 x\, \sqrt{-g} \left[ \frac{m_{Pl}^2}{16\pi } R + \frac{1}{M^2} (\partial_{\mu} R) J^{\mu}  - 
g^{\mu \nu} \partial_{\mu}\phi\, \partial_{\nu}\phi^* + U(\phi, \phi^*)\right] - A_m\, ,
\label{act-tot}
\end{eqnarray}
where $m_{Pl}=1.22\cdot 10^{19}$ GeV is the Planck mass, 
$A_m$ is the matter action,  $J^\mu = g^{\mu\nu}J_\nu$, and $g^{\mu\nu}$ is the metric tensor of the 
background space-time. We assume that initially the metric has the usual GR form and study the emergence 
of the corrections due to the  instability described below.

In contrast to scalar electrodynamics, the baryonic current of scalars is not uniquely defined. In electrodynamics 
the form of the electric current  is dictated by the conditions of gauge invariance and current conservation, which
demand the addition to the current % (\ref{current}})
of the so called sea-gull term proportional to $e^2 A_\mu |\phi|^2$, where $A_\mu$ is the electromagnetic potential.

On the other hand, a local $U(1)$-symmetry is not imposed on the theory determined by action (\ref{act-tot}).
It is invariant only with respect to  a $U(1)$ transformations with a constant phase.
As a result,  the baryonic current of scalars
is considerably less restricted. In particular, we can add to the current an analogue of the sea-gull term, 
$ \sim (\partial_\mu R)\,|\phi |^2 $, with an arbitrary coefficient.

In our paper~\cite{Arbuzova:2016cem} we study the following two extreme possibilities, when the sea-gull term is absent 
and the current is not conserved, or the sea-gull term is included with the coefficient ensuring current conservation.
%with the coefficient which ensures current conservation in curved space-time with curvature $R$ .
 In both cases no baryon asymmetry can be generated without additional interactions. 
It is trivially true in the second case, when the current is conserved,
but it is also true in the first case despite 
the current non-conservation, simply because the non-zero divergence $D_\mu J^\mu$
does not change the baryonic number of $\phi$ but only leads to redistribution of particles $\phi$ in the phase space.
So to create any non-zero baryon asymmetry we have to introduce an interaction of $\phi$ with other particles 
which breaks  conservation of $B$ by making the potential $U$ non-invariant with respect to 
the phase rotations of $\phi$, as it is described below.

If the potential $U(\phi)$
is not invariant with respect to the $U(1)$-rotation, $\phi \rightarrow \exp{(i \beta)} \phi $, 
the baryonic current  defined in the usual way    
\begin{eqnarray}
J_{ \mu} = i q (\phi^* \partial_{\mu}\phi - \phi \partial_{\mu}\phi^*)
\label{current}
\end{eqnarray}
is not conserved. Here $q$ is the baryonic number of $\phi $ and
 we omitted index $B$ in current $J_{\mu}$.  

With this current and Lagrangian (\ref{act-tot}) the equation  
for the curvature scalar, $R$, takes the form:
\begin{eqnarray}
\frac{m_{Pl}^2}{16\pi }\, R + \frac{1}{M^2}\left[ (R + 3 D^2) D_{\alpha} J^{ \alpha} + J^{\alpha} \,D_{\alpha}R  \right] - 
D_{\alpha} \phi \, D^{\alpha} \phi^* + 2 U(\phi ) = - \frac{1}{2} \, T_{\mu}^{\mu}\, ,
\label{trace-eq}
\end{eqnarray}
where $D_\mu$ is the covariant derivative in metric $g_{\mu\nu}$ (of course, for scalars $D_\mu = \partial_\mu$) and
$T_{\mu\nu}$ is the energy-momentum tensor of matter obtained from action $A_m$.  

According to definition (\ref{current}), the current divergence is:
\begin{eqnarray}
D_{\mu} J^{\mu} =  \frac{2q^2}{M^2} \left[ D_{\mu} R\, (\phi^* D^{\mu}\phi + \phi D^{\mu}\phi^*) + |\phi|^2 D^2 R \right]
+ i q \left(\phi \frac{\partial U}{\partial \phi} - \phi^* \frac{\partial U}{\partial \phi^*} \right)\,.
\label{J-div}
\end{eqnarray}
If the potential of $U$ is invariant with respect to the phase rotation of $\phi$, i.e.
$U = U(|\phi|)$, the last term in this expression disappears. Still the current remains non-conserved, but this non-conservation
does not lead to any cosmological baryon asymmetry. Indeed, the current non-conservation is proportional to the product
$\phi^* \phi$, so it can produce or annihilate an equal number of baryons and antibaryons.

To create cosmological baryon asymmetry we need to introduce new types of interactions,
for example, the term in the potential of the form:
$ U_4 = \lambda_4 \phi^4 + \lambda_4^* \phi^{*4} $. This potential is surely non invariant w.r.t. the phase rotation of $\phi$
and can induce the B-non-conserving process of transition of two scalar baryons into two antibaryons,
$2 \phi \rar 2 \bar \phi$.

Let us consider solution of the above equation of motion in cosmology. The metric of the 
spatially flat cosmological FRW background can be taken as:
\be
ds^2=dt^2 - a^2(t) d{\bf r}^2\, .
\label{ds-2}
\ee

In the homogeneous case the  equation for the curvature scalar (\ref{trace-eq}) takes the form:
\be 
\frac{m_{Pl}^2}{16\pi }\, R + \frac{1}{M^2}\left[ (R + 3 \partial_t^2 + 9 H \partial_t) D_{\alpha} J^{\alpha} + 
\dot R \, J^0 \right] 
%\dot \phi \, \dot \phi^* - 2 U(\phi ) 
= - \frac{ T^{(tot)}}{2} \,, 
\label{trace-eq-FRW}
\ee
where  $J^0$ is the baryonic number density of the $\phi$-field,
$H = \dot a/a$ is the Hubble parameter, and $T^{(tot)}$ is the trace of the energy-momentum tensor of matter including 
contribution from the $\phi$-field. In the homogeneous and isotropic cosmological plasma 
\be
T^{(tot)} = \rho - 3 P\, ,
\label{T-tot}
\ee
where $\rho$ and $P$ are respectively the energy density and the pressure of the plasma. 
For relativistic plasma $\rho = \pi^2 g_* T^4/30$ with $T$ and $g_*$ 
being the plasma temperature and the number of particle species in the plasma. 
The Hubble parameter is expressed through $\rho$ as $H^2 = 8\pi \rho/(3m_{Pl}^2) \sim T^4/m_{Pl}^2$. 

%In equation (\ref{trace-eq-FRW}) $J_1^0$ is the density of the baryonic number of the $\phi$-field and 
The covariant divergence of the current is given by the expression (\ref{J-div}). In the
homogeneous case we are considering it takes the form:
\be 
D_{\alpha} J^{\alpha} = \frac{2q^2}{M^2} \left[ \dot R\, (\phi^* \dot \phi + \phi \dot \phi^*) + 
(\ddot R + 3H \dot R)\, \phi^*\phi\right]
+ i q \left(\phi \frac{\partial U}{\partial \phi} - \phi^* \frac{\partial U}{\partial \phi^*} \right)\,.
\label{J-div-hom}
\ee

To derive the equation of motion for the classical field $R$ in the cosmological plasma we have to take 
the expectation values of the products of the quantum operators $\phi$, $\phi^*$, and their derivatives. 
Performing  the thermal averaging, we find 
\be
\langle \phi^* \phi \rangle =  \frac{T^2}{12}\, , \ \ \ 
\langle \phi^* \dot \phi + \dot \phi^* \phi \rangle = 0\, .
\label{av-phi}
\ee
Substituting these average values into Eq. (\ref{trace-eq-FRW}) and neglecting the last term in 
Eq.~(\ref{J-div-hom}) we obtain the fourth order differential equation:
\be
\frac{m_{Pl}^2}{16\pi }\, R + 
\frac{q^2}{6 M^4} \left(R + 3 \partial_t^2 + 9 H \partial_t \right)
\left[ 
\left(\ddot R + 3H  \dot R\right) T^2 \right] + 
\frac{1}{M^2} \dot R \, \langle J^0 \rangle 
= - \frac{T^{(tot)}}{2} \, .
\label{trace-eq-plasma}
\ee
Here $\langle J^0 \rangle $ is the thermal average value of the baryonic number density of $\phi$. It is assumed 
to be zero initially  and generated as a result of GBG. We neglect this term, since it is surely small initially and
probably subdominant later. Anyhow it does not noticeably change the exponential rise of $R$ at the onset of the
instability.

Eq.~(\ref{trace-eq-plasma}) can be further simplified if the variation of $R(t)$ is much faster than the universe expansion 
rate or in other words $\ddot R / \dot R \gg H$. Correspondingly the temperature may be considered adiabatically
constant. The validity of these assumption is justified a posteriori after we find  the solution for $R(t)$. 

Keeping only the linear in $R$ terms and neglecting higher powers of $R$, such as $R^2$ or $H R$, we obtain the
linear differential equation of the fourth order:
\be
\frac{d^4 R}{dt^4} + \mu^4 R =  - \frac{1}{2} \, T^{(tot)}\,,\  {\rm where }\ \   \mu^4 = \frac{m_{Pl}^2 M^4}{8 \pi q^2 T^2}\,.
\label{d4-R}
\ee

The homogeneous part of this equation has exponential solutions  $R \sim \exp (\lambda t)$ with
\be
\lambda = | \mu | \exp \left(  i\pi /4 + i \pi n /2 \right),
\label{lambda}
\ee
where $n = 0,1,2,3$. 

There are two solutions with positive real parts of $\lambda$.
This indicates that the curvature scalar is
exponentially unstable with respect to small perturbations, so $R$ should rise exponentially fast with time 
and quickly oscillate around this rising function.

Now we need to check if the characteristic rate of  the perturbation explosion is indeed much larger than the rate
of the universe expansion, that is:
\be 
(Re\, \lambda)^4   > H^4 = \left( \frac{ 8\pi \rho }{ 3 m^2_{Pl}}\right)^2 =  \frac{16 \pi^6 g_*^2}{2025}\,
\frac{ T^8}{m_{Pl}^4}, 
\label{lambda-to-H}
\ee
where $\rho = \pi^2 g_* T^4 /30$ is the energy density of the primeval plasma at temperature $T$
and $g_* \sim 10 - 100$ is the number of relativistic
degrees of freedom in the plasma. This condition is fulfilled if
\be
    \frac{2025}{2^9 \pi^7 q^2 g_*^2}\frac{m_{Pl}^6 M^4}{T^{10}} > 1\,,
\label{inst-OK}
\ee
or, roughly speaking, if $T \leq m_{Pl}^{3/5} M^{2/5} $.  Let us stress that at these temperatures the
instability is quickly developed and the standard cosmology would be destroyed.

If we want to preserve the successful  big bang nucleosynthesis (BBN)  results and  
impose the condition that the development of the instability 
was longer than the Hubble time at the BBN epoch at $T \sim 1 $ MeV, then $M$ should be extremely small,
$M < 10^{-32}$ MeV. The desire to keep the standard cosmology at smaller $T$ would demand even tinier $M$.
A tiny $M$ leads to a huge strength of coupling (\ref{L-GBG}). It surely would lead to pronounced effects in
stellar physics. 

\section{Gravitational baryogenesis with fermions \label{s-fermions}}

Let us now generalize results, obtained for scalar baryons, to realistic fermions. We start from the action in the form
\begin{eqnarray}
A= \int d^4x \sqrt{-g} \left[\frac{m_{Pl}^2}{16 \pi} \,R - {\cal L}_{m}\right]\, 
\label{A-gen}
\end{eqnarray}
with  
\begin{eqnarray} \nonumber
{\cal L}_{m} &= &\frac{i}{2} (\bar Q \gamma ^\mu \nabla _\mu Q -   \nabla _\mu \bar Q\, \gamma ^\mu Q) - m_Q\bar Q\,Q\\ 
&+& 
\frac{i}{2} (\bar L \gamma ^\mu \nabla _\mu L -   \nabla _\mu \bar L \gamma ^\mu L)
- m_L\bar L\,L \\ \nonumber
&+& \frac{g}{m_X^2}\left[(\bar Q\,Q^c)(\bar Q L) + (\bar Q^cQ)(\bar L Q) \right]
+ \frac{f}{m_0^2} (\partial_{\mu} R) J^{\mu} + {\cal L}_{other}\,, 
\label{L-matt}
\end{eqnarray}
where $Q$ is the quark (or quark-like) field with non-zero baryonic number, $L$ is
another fermionic field (lepton),
$\nabla_\mu  $ is the covariant derivative of Dirac fermion in tetrad formalism. $m_0$ is a constant parameter with dimension of mass and $f$ is dimensionless
coupling constant which is introduced to allow for an arbitrary sign of the curvature dependent term in the above expression. 
$J^{\mu } = \bar Q \gamma ^{\mu} Q$ is the quark current with $\gamma ^{\mu}$ being the curved space gamma-matrices,  
${\cal L}_{other}$ describes all other forms of matter.
The four-fermion interaction between quarks and leptons is introduced to ensure the
necessary non-conservation of the baryon number with $m_X$ being a constant
parameter with dimension of mass and $g$ being a dimensionless coupling constant.
In grand unified theories $m_X$ may be of the order of $10^{14}-10^{15}$ GeV.

Varying  the action (\ref{A-gen}) over metric, $g^{\mu \nu}$, and taking trace with respect to $\mu$ and $\nu$, we obtain the following equation of motion for the curvature scalar:
\begin{eqnarray} \nonumber
- \frac{m_{Pl}^2}{8\pi } R &=& m_{Q} \bar Q Q + m_L \bar L L 
+
 \frac{2g}{m_X^2}\left[(\bar Q\,Q^c)(\bar Q L) + (\bar Q^cQ)(\bar L Q) \right] \\
 &-& 
 \frac{2f}{m_0^2} (R + 3D^2) D_{\alpha} J^{\alpha} + T_{other}\,,
 \label{trace}
 \end{eqnarray}
where $T_{other} $ is the trace of the energy momentum tensor of all other fields.
At relativistic stage, when masses are negligible, we can take  $T_{matter} = 0$. 
The average  expectation value of the interaction term proportional to $g$ is also 
small, so the contribution of all matter fields may be neglected.

As we see in what follows, kinetic equation leads to an explicit dependence on $R$
of the current divergence, $D_\alpha J^\alpha$, if the current is not conserved. As
a result we obtain 4th order equation for $R$.

As previously, we study solutions of Eq.~(\ref{trace}) in cosmology in homogeneous and
isotropic FRW background with the metric $ds^2=dt^2 - a^2(t) d{\bf r}^2 $.
The curvature is a function only of time and the covariant derivative acting on a vector
$V^\alpha$, which depends only on time and has only time component, has the form:
\begin{eqnarray}
D_\alpha V^\alpha = (\partial_t + 3 H) V^t ,
\label{DV}
\end{eqnarray}
where $H = \dot a/a$ is the Hubble parameter.

As an example let us consider the reaction $q_1 + q_2 \leftrightarrow \bar q_3 + l_4$, where $q_1$ and
$q_2$  are quarks with momenta $q_1$ and $q_2$, while $\bar q_3$ and $l_4$ are antiquark
and lepton with momenta $q_3$ and $l_4$. We use the same notations for the particle
symbol and for the particle momentum.
The kinetic equation for the variation of the baryonic number density $n_B \equiv J^t$
through this reaction in the FRW background has the form: 
\begin{eqnarray}
(\partial_t +3 H) n_B = I_B^{coll}, 
\label{kin-eq-gnrl}
\end{eqnarray}
where the collision integral for space and time independent interaction is equal to:
\begin{eqnarray} 
&&I^{coll}_B =- 3 B_q (2\pi)^4  \int  \,d\nu_{q_1,q_2}  \,d\nu_{\bar q_3, l_4}
\delta^4 (q_1 +q_2 -q_3 - l_4)
\nonumber\\
&& \left[ |A( q_1+q_2\rightarrow  \bar q_3 +l_4)|^2
f_{q_1} f_{q_2} -
 |A( \bar q_3 +l_4  \rightarrow  q_1+q_2 ) |^2
 f_{\bar q_3} f_{l_4}
 \right],
\label{I-coll}
\end{eqnarray}
where $ A( a \rightarrow b)$ is the amplitude of the transition from state $a$ to state $b$,
$B_q$ is the baryonic number of quark, $f_a$ is the phase space distribution (the
occupation number), and 
\begin{eqnarray}
d\nu_{q_1,q_2}  =  
\frac{d^3 q_1}{2E_{q_1} (2\pi )^3 }\,  \frac{d^3 q_2}{2E_{q_2} (2\pi )^3 } ,
\label{dnuy}
\end{eqnarray}
where $E_q = \sqrt{ q^2 + m^2}$ is the energy of particle with three-momentum
$q$ and mass $m$. The element of phase space of final particles, $d\nu_{\bar q_3, l_4} $, is defined analogously. 

We neglect the Fermi suppression factors and the effects of gravity
 in the collision integral. This is generally a good approximation.

The calculations are strongly simplified if quarks and leptons are in equilibrium
with respect to elastic scattering and annihilation. In this case their distribution functions 
take the form
\begin{eqnarray}
f = \frac{1}{e^{(E/T - \xi}) + 1} \approx e^{-E/T + \xi}, 
\label{f-eq}
\end{eqnarray}
with $\xi = \mu/T$ being dimensionless chemical potential, different for quarks, $\xi_q$,
and leptons, $\xi_l$. 

The assumption of kinetic equilibrium is well justified since it is usually enforced by 
very efficient elastic scattering. Equilibrium with respect to annihilation, say, 
into two channels: $2\gamma$ and $3\gamma$, implies 
the usual relation between chemical potentials of particles and antiparticles,
$\bar \mu = -\mu$. 

The baryonic number density is given by the expression:
\begin{eqnarray} \nonumber
n_B &=& \int \frac{d^3 q}{2 E_q\, (2\pi)^3}  (f_q - f_{\bar q})  \\
&=& \frac{g_S B_q}{6} \left(\mu T^2 + \frac{\mu^3}{ \pi^2}\right) =
\frac{g_S B_q T^3}{6}\,\left(\xi + \frac{\xi^3}{\pi^2}\right) \,,
\label{n-B-of-mu}
\end{eqnarray}
where  $T$ is the cosmological plasma temperature, $g_S$ and $B_q$ are respectively
 the number of the spin states and the baryonic number of quarks. 
 
We can use another  representation of the quark field:
\begin{eqnarray} 
Q_2 = \exp( i f R /m_0^2 )\, Q
\label{Q2}
\end{eqnarray}
analogously to what is done in our paper \cite{Arbuzova:2016qfh}.
Written in terms of $Q_2$ Lagrangian (\ref{L-matt}) would not contain terms proportional
to $f /m_0^2$, but dependence on such terms would reappear in the interaction term
as:
\begin{eqnarray}
 \frac{2g}{m_X^2}\left[ e^{-3ifR/m_0^2}\, (\bar Q_2\,Q_2^c)(\bar Q_2 L) + 
 e^{3ifR/m_0^2}\,(\bar Q_2^cQ_2)(\bar L Q_2) \right] .
\label{int-R}
\end{eqnarray}
Nevertheless we obtain the same fourth order equation for the evolution of curvature,
as for non-rotated field $Q$.

 Since the transition amplitudes, which enter the collision integral, are obtained by
integration over time of the Lagrangian operator (\ref{int-R}), taken
between the initial and final states, the energy conservation delta-function in 
Eq. (\ref{I-coll}) would be modified due to time dependent factors $\exp [\pm 3 ifR(t)/m_0^2]$. 
In the simplest case, which is usually considered
in gravitational (and spontaneous) baryogenesis, a slowly
changing $\dot R$ is taken, so we can approximate $R(t) \approx \dot R(t)\,t$.  
In this case the energy is not conserved but the energy
conservation condition is trivially modified, as 
\begin{eqnarray} \nonumber
&&\delta [ E(q_1) + E(q_2) - E(q_3) - E(l_4) ] \rightarrow \\
&&\rightarrow \delta [ E(q_1) + E(q_2) - E(q_3) - E(l_4) - 3f\dot R(t) /m_0^2\,]. 
\label{E-non-conserved}
\end{eqnarray}
Thus the energy is non-conserved due to the action of the external field $R(t)$.
Delta-function (\ref{E-non-conserved}) is not precise, but the result is pretty close to it,  
 if $\dot R(t)$    changes very little
during the effective time of the relevant reactions.

% In the simplest case, which is usually considered
%in gravitational (and spontaneous) baryogenesis, a slowly
%changing $\dot R$ is taken, so we can approximate $R(t) \approx \dot R(t)\,t$.  
If the dimensionless chemical potentials $\xi_q$ and $\xi_l$, as well as 
$ f\dot R(t) /m_0^2 /T $, are small, 
the collision integral can be written as:
\begin{eqnarray}
I^{coll}_B \approx \frac{C_I g^2 T^8}{m_X^4}\,  
\left[ \frac{3 f\dot R(t)}{ m_0^2\,T} - 3\xi_q + \xi_l  \right] , 
\label{Icoll-appr}
\end{eqnarray}
where $C_I$ is a positive dimensionless constant. 
The factor $T^8$ appears for reactions with massless particles and 
the power eight is found from dimensional consideration. 
Because of conservation of
the sum of baryonic and leptonic numbers $\xi_l = -\xi_q/3 $.

The case of an essential variation of $\dot R (t)$ is analogous to fast
variation of $\dot \theta (t)$ studied in our paper~\cite{Arbuzova:2016qfh}. 
Clearly, it is  much more complicated technically. Here we consider only the simple 
situation with quasi-stationary background and postpone more realistic time
dependence of $R(t)$ for the future work.

For small chemical potential the baryonic number density (\ref{n-B-of-mu}) is equal to 
\begin{eqnarray}
n_B \approx \frac{g_s B_q}{6}\, \xi_q T^3\, ,
\label{n-small-xi}
\end{eqnarray}
and if the temperature adiabatically decreases
in the course of the cosmological expansion, according to $\dot T = - H T$, equation
(\ref{kin-eq-gnrl}) turns into
\begin{eqnarray}
\dot \xi_q = \Gamma \left[ \frac{9 f\dot R(t)} {10m_0^2\, T} - \xi_q \right],
\label{dot-xi-q}
\end{eqnarray}
where $\Gamma \sim g^2 T^5/m_X^4 $ is the rate of B-nonconserving reactions.

If $\Gamma$ is in a certain sense large, this equation can be solved in stationary
point approximation as
\begin{eqnarray} 
\xi_q=\xi_q^{eq} - \dot \xi_q^{eq}/\Gamma\, , \ \ {\rm where} \ \ \xi_q^{eq} = \frac{9}{10} \frac{f \dot R} {m_0^2 T}\, .
\label{xi-approx}
\end{eqnarray}
If we substitute $\xi_q^{eq} $  into Eq.~(\ref{trace}) we arrive 
to the fourth order equation for $R$.

According to the comment below Eq. (\ref{trace}), the contribution of thermal matter
into this equation can be neglected, and we arrive to the very simple fourth order
differential equation:
\begin{eqnarray}
\frac{d^4 R}{dt^4} = \lambda^4 R,
\label{d4-R}
\end{eqnarray}
where $\lambda^4 = C_\lambda m_{Pl}^2 m_0^4 /T^2 $ with
$ C_\lambda  = 5/(36\pi f^2 g_s B_q )$. Deriving this equation we neglected
the Hubble parameter factor in comparison with time derivatives of $R$. It is justified a posteriori because
the calculated $\lambda$ is much larger than $ H $.

Evidently equation (\ref{d4-R}) has extremely unstable solution with instability time
by far shorter than the cosmological time. This instability would lead to an explosive
rise of $R$, which may possibly be terminated  by the nonlinear  terms proportional to the
product of $H$ to lower derivatives of $R$. Correspondingly one may expect stabilization when $HR \sim \dot R$,
i.e. $H\sim \lambda$. Since 
\begin{eqnarray}
\dot H + 2 H^2 = - R/6,
\label{dot-H}
\end{eqnarray}
$H$ would also exponentially  rise together with  $R$, 
$ H \sim \exp (\lambda t )$ and $\lambda H \sim R$.
Thus stabilization may take place at $R \sim \lambda^2   \sim  m_{Pl} m_0^2 / T$.
This result  should be compared with the normal General Relativity value
$ R_{GR} \sim T_{matter} /m_{Pl}^2 $, where $T_{matter}$ is the trace of the 
energy-momentum tensor of matter.

\section{Discussion and conclusion \label{sec-concl}}

For more accurate analysis numerical solution will
be helpful, which we will perform in another work. The problem is complicated because
the assumption of slow variation of $\dot R$ quickly becomes broken and the
collision integral in time dependent background is not so simply tractable as 
the usual stationary one.  The technique for treating kinetic equation in non-stationary
background is presented  in Ref.~\cite{Arbuzova:2016qfh}. For evaluation of $R(t)$ in this case
numerical calculations are necessary, which will be presented elsewhere. Here 
we describe only the basic features of the new effect of instability in gravitational baryogenesis.

To conclude we have shown that gravitational baryogenesis in the simplest versions
discussed in the literature is not realistic because the instability
of the emerging gravitational  equations destroys the standard cosmology. Some
stabilization mechanism is strongly desirable. Probably stabilization may be achieved in a version
of $F(R)$-theory.

\vspace{1cm}
\centerline{{\bf Acknowledgement}}

This work was supported by the RSF Grant N 16-12-10037.
The author expresses sincere gratitude to Harald Fritzsch for his invitation and for the opportunity to present the talk at the Conference on Particles and Cosmology. 
She would like to thank  Kok Khoo Phua for his kind hospitality at NTU, Singapore. 

%Non BiBTeX users can list down their references as:

\end{document}